%% file: main.tex
\pdfoutput=1

\documentclass[11pt]{article}

\usepackage[final]{acl}
\usepackage{times}
\usepackage{latexsym}

\usepackage[T1]{fontenc}

\usepackage[utf8]{inputenc}

\usepackage{microtype}

\usepackage{inconsolata}

\usepackage{graphicx}

\usepackage{svg}
\usepackage{array}
\usepackage{adjustbox}
\usepackage{caption}
\usepackage{multirow}
\usepackage{booktabs}
\usepackage{listings}
\usepackage{subcaption}
\usepackage{bbding}
\usepackage{cleveref}
\usepackage{xfrac}
\usepackage{svg}
\usepackage{float}
\usepackage{xcolor}
\usepackage{hyperref}
\usepackage{amssymb}

\usepackage[T1]{fontenc}

\newcommand{\pisa}{\texttt{PISA2015}}
\newcommand{\interalia}{\emph{inter alia}}
\newcommand{\hai}{human-AI}
\newcommand{\cpsrp}{\texttt{CPS-TaskForge}}
\newcommand{\cpscheck}{\texttt{CPS-\checkmark}}

\newcommand\uw{$^\spadesuit$}

\newcommand\ucsd{$^{\clubsuit}$}
\newcommand\ai{$^{\diamondsuit}$}
\newcommand\aspace{\hspace{.75em}}

\graphicspath{{./src/img/}}

\title{\cpsrp: Generating 
Collaborative Problem Solving Environments for Diverse Communication Tasks}

\author{
  Nikita Haduong\uw\aspace
  Irene Wang\aspace
  Bo-Ru Lu\uw\aspace \\
  \textbf{Prithviraj Ammanabrolu}\ucsd\aspace
  \textbf{Noah A. Smith}\uw\ai\aspace
  \\
  \uw{}University of Washington \aspace
  \ucsd{}University of California, San Diego \aspace
  \ai{}Allen Institute for AI\\
  {\tt \{qu,nasmith\}@cs.washington.edu roylu@washington.edu
  }
  {\tt 
  prithvi@ucsd.edu
  }
}

\begin{document}

\maketitle

\input{src/0_abstract_v4}

\newcommand{\collabps}{$C_{ollab}PS$}
\newcommand{\coopps}{$C_{oop}PS$}

\input{src/1_intro_v4}
\input{src/2_background}

\input{src/3_research_platform_v2}

\input{src/10_cps_checklist}
\input{src/7_case_study}

\input{src/4_analysis}

\input{src/11_related_work}

\input{src/6_discussion}

\input{src/z_extras}

% \bibliography{custom}

\input{main.bbl}
\appendix 
\input{src/x_appendix.tex}

\end{document}

%% file: src/0_abstract_v4.tex
\begin{abstract}
    Teams can outperform individuals; could adding AI teammates further bolster performance of teams solving problems collaboratively? 
    Collaborative problem solving (CPS) research commonly studies teams with two agents (human-human or human-AI), but team research literature finds that, for complex tasks, larger teams are more effective. 
    Progress in studying collaboration with more than two agents, through textual records of team interactions, is hindered by a major data challenge: available CPS corpora are predominantly dyadic, and adapting pre-existing CPS tasks to more agents is non-trivial.
    We address this data challenge by developing a CPS task generator, \cpsrp, that can produce environments for studying CPS under a wide array of conditions, and releasing a CPS task design checklist grounded in the theoretical PISA 2015 CPS framework to help facilitate the development of CPS corpora with more agents.
    \cpsrp{} takes the form of a resource management (tower defense) game, and different CPS tasks can be studied by manipulating game design parameters.
    We conduct a case study with groups of 3--4 humans to validate production of diverse natural language CPS communication in a game instance produced by \cpsrp.
    We discuss opportunities for advancing research in CPS (both with human-only and human-AI teams) using different task configurations.
    We will release data and code.\footnote{\url{https://github.com/nhaduong/cps-taskforge}}
\end{abstract}

%% file: src/1_intro_v4.tex
\section{Introduction}

Modern life requires teamwork to solve problems~\citep{marks2001}, but what makes a team work well together? This area of study, known as collaborative problem solving (CPS), is active across many disciplines, e.g., psychologists study the construction of team mental models in team discussions~\cite{Lee2015AnalysisOT}, business management sciences investigate how communication style affects performance evaluation~\citep{10.2308/TAR-2020-0198}, and educators develop tools to teach team communication strategies~\citep{CPSCoach}, emphasizing the research direction of discovering \emph{how team members talk to one another}. 
Conducting empirical work in CPS faces many challenges, in large part because of a large CPS task design space (e.g., what is the problem, who makes up the team, and who knows what information when). 
As a result, despite extensive interdisciplinary work in CPS,  task designs in empirical studies have often focused on teams of two collaborating to solve  problems such as selecting a designated object, modeling search and rescue, and making decisions.

AI agents have the potential to increase team effectiveness, and developing ways to integrate AI into teams is an active area of research in communities such as HCI~\citep{Cai2019HelloAU}, NLP~\cite{Bansal2019BeyondAT,vats2024survey}, and AI fairness~\citep{Lai2021TowardsAS}. 
Example integrations include AI-assisted decision making with one human and one AI (e.g., cancer diagnosis, \citealp{https://doi.org/10.1002/cac2.12215}) and AI-assisted creative tooling (e.g.,~\citealp{Tsiros2020TowardsAH,lu2024does}). 
Developing these collaborative tools is made possible through open datasets. For example, various Amazon reviews datasets (e.g., \citealp{fornaciari-poesio-2014-identifying} and \citealp{ni-etal-2019-justifying}) have been used to develop sentiment classifiers and deception detectors that can be used as AI-assisted decision makers, 
and the Reddit WritingPrompts dataset~\cite{fan-etal-2018-hierarchical} has been valuable in developing co-writing AI systems. 
Unfortunately, a paucity of open datasets with more than two parties leads to challenges in integrating AI with larger human teams, as we lack understanding of team dynamics when an AI communicates to a team, rather than an individual.

To support CPS study across different designs (e.g., adding a third AI teammate to a two-human team or using voice instead of text communication), we introduce a CPS task environment generator, \cpsrp{}.
\cpsrp{} instantiates a \emph{resource management} activity through a \textbf{tower defense} game and supports adjusting a range of CPS design parameters such as team composition, communication method, and how stressful the task is. 
In a tower defense game, players must defend their base by using limited resources to construct towers that can defeat enemies before the enemies destroy the base.
We provide a CPS task design checklist, \cpscheck, adapted from the PISA 2015 theoretical CPS framework (\pisa) developed by the OECD~\citep{pisa-2015}, to support generating the desired task environment with \cpsrp.

We illustrate \cpsrp{} capabilities by presenting several CPS task designs and conducting a case study that can collect human communication data exhibiting a range of CPS skills, including social skills such as maintaining group communication and cognitive skills such as developing strategic plans. Our study has small groups of 3--4 participants complete a task multiple times with increasing difficulty.
We observe many different successful strategies and a wide range in CPS skill usage across teams, demonstrating the versatility of collecting data through \cpsrp.

To summarize our contributions:

\begin{enumerate}
    \item We identify opportunities and gaps in the interdisciplinary CPS literature. We argue that human team research can help advance human-AI team design; however, there exist challenges associated with the lack of diverse CPS data available to the research community.
    \item We introduce \cpsrp{}, which allows researchers to generate a variety of CPS task environments for studying human and human-AI CPS team processes. We adapt a theoretical CPS framework into a design checklist, \cpscheck, to assist with \cpsrp{} environment generation.
    \item We present a case study using \cpsrp{} to illustrate the variability of CPS data through a study with more than two agents.
    We release the conversation and game interaction data collected during the study as an example of what can be produced using \cpsrp.
\end{enumerate}

%% file: src/2_background.tex
\section{Collaboration and Problem Solving}
\input{src/tab/data_stats}

Collaborative problem solving (CPS) processes are well-studied for human teams, but when \hai{} teams are considered,  downstream task performance has been prioritized, leaving \hai{} CPS processes understudied. 
For example, ~\citet{10.2308/TAR-2020-0198} found human team communication more effective when the appropriate style was used in conjunction with the delivery of relevant information. 
Humans have different expectations towards AI teammates \citep{zhang2023trust,zhang2021ideal,GRIMES2021113515}, so \hai{} teams may value communication style differently. 
Studying \hai{} CPS processes requires developing the appropriate datasets, but resources for creating such data is deficient.

Understanding how effective and efficient communication can predict successful teamwork requires collecting data in a variety of CPS settings. The tasks used to elicit relevant data often model real-world activities, e.g., rescuing humans from a burning building (ASIST; \citealp{corral2021building,freeman2021evaluating}), instruction following through selecting designated objects (e.g., PentoRef,~\citealp{zarriess-etal-2016-pentoref}; KTH Tangrams,~\citealp{shore-etal-2018-kth}; PhotoBook,~\citealp{takmaz-etal-2020-refer-photobook}; Doll Dialogue, \citealp{dolldialogue}; \citealp{bloco-building}), and navigating environments (e.g., HCRC Map Task,~ \citealp{anderson1991hcrc}; \citealp{effenberger-etal-2021-analysis-language}), and use human participants.
The resulting datasets have been used to study a wide variety of communication and linguistic phenomena, including language entrainment (i.e., when communicative behavior becomes similar among interlocutors, including lexical choice and rhythm) and common ground building (i.e., when interlocutors develop their own code).
To the best of our knowledge, analogous settings incorporating an AI team member in a CPS task have not explored similar communication and linguistic phenomena because only recently has AI-generated natural language become indistinguishable from humans~\citep{clark-etal-2021-thats,Dugan2022RealOF}, enabling exploration of AI teammates as peers.
Unfortunately, expanding pre-existing datasets to other CPS settings, such as involving an AI agent or a third human team member, is challenging because the tasks were designed to study a specific team composition; for example, what role would a third participant play in a navigation task originally designed for one human to tell another human where to go?

Despite the extensive body of literature studying CPS, publicly available resources remain scarce, particularly when more than two agents are involved.
We summarize a sample of CPS task activities in the literature in \autoref{tab:dataset_stats} to illustrate gaps in task type and team size between studies with or without data release to the research community.

%% file: src/tab/data_stats.tex
\begin{table*}[!htp]\centering
	\scriptsize

	\begin{tabular}{lcrc}\toprule
		                                                                & Task type                       & Team Size & Communication Modality \\\midrule %
		KTH Tangrams~\citep{shore-etal-2018-kth}                        & Object Identification                & 2         & Speech                 \\ %
		PentoRef~\citep{zarriess-etal-2016-pentoref}                    & Object Identification                & 2         & Multimodal             \\ %
		TEAMS~\citep{rockenbach2007teams}                               & Forbidden Island~\texttrademark & 3--4      & Multimodal             \\ %
		ASIST~\citep{ASU/BZUZDE_2022}                                   & Search and Rescue               & 3         & Multimodal             \\ %
		CerealBar~\citep{suhr-etal-2019-executing}                      & Search and Rescue               & 2         & Text                   \\ %
		HCRC Map Task~\citep{anderson1991hcrc}                          & Search and Rescue                      & 2         & Speech                 \\ %
		PhotoBook~\citep{takmaz-etal-2020-refer-photobook}              & Object Identification                & 2         & Text                  \\ %
		Cards~\citep{Potts:2012-cards}                                  & Search and Rescue               & 2         & Text                   \\ %
		\midrule %
		\citet{friendship}                                              & Object Identification                & 2         & Multimodal             \\ %
		\citet{ma_2023_7304816}                                         & Programming                     & 2         & Multimodal             \\ %
		\citet{block-collection}                                        & Search and Deliver              & 2         & Text                   \\ %
		\citet{Kokel2022HumanguidedCP}                                  & Object Construction             & 2         & Multimodal             \\ %
		\textcolor{red}{$^\bullet$} MRE \citep{Hill2003VirtualHI}                        & Decision Making                 & 21        & Speech                 \\ %
		T-shirt Task \cite{cps-math}                                         & Math Problem                  & 2         & Multimodal             \\ %
		Volcano Lab \cite{flor-etal-2016-automated}                      & Science Lab                    & 2         & Text                   \\ %
		Circuit Lab \cite{circuit}                                      & Science Lab                 & 3         & Text                   \\ %
		Physics Playground~\citep{generalized-cps}      & 2D Physics Puzzles              & 3         & Multimodal             \\ %
		Minecraft~\citep{generalized-cps}    & Minecraft Hour of Code          & 3         & Multimodal             \\ %
		CPSCoach~\citep{CPSCoach}                                       & 2D Physics Puzzles              & 2         & Multimodal             \\ %
		\textcolor{red}{$^\bullet$} NeoCities \cite{10.1145/3492832}                    & Search and Rescue               & 3         &  Text                      \\ %
		9-11 Firefighting \citep{hutchins2008understanding}					& Firefighting 					&	---			& Speech					\\
		Air Warfare	\citep{hutchins2008understanding}														& Object Identification				& 6+		& Speech				\\		%
		Maritime Interdiction Operations	\citep{hutchins2008understanding}								& Object identification				& 3+		& Speech \\
		\citet{wiltshire2018problem}	&	NASA Moonbase Alpha Simulation	& 2	& Speech \\
		\midrule
		\cpsrp{} (this work)							&	Object Identification, Resource Management	&	1--4+	&	Text, Speech \\
		\bottomrule
	\end{tabular}
 	\caption{A sample of collaborative problem solving research. The top group contains work that produced datasets open to the research community. \textcolor{red}{$^\bullet$} indicates studies with AI teammates. 
	Object identification tasks require identifying an object, search and rescue requires navigating an environment to locate an object, and search and deliver requires returning to a second point after locating the object. The math and science lab tasks are typical tasks found in educational contexts.
	Forbidden Island\texttrademark{} is a commercial cooperative board game.
	``Text'' data often contains system interaction log data such as mouse clicks, whereas ``Multimodal'' communication may include video of participant bodies, audio, and hormonal measurements.
	We observe more diverse tasks conducted in works without open data. 
	}
	\label{tab:dataset_stats}
\end{table*}

%% file: src/3_research_platform_v2.tex
\section{\cpsrp{} and Tower Defense}

\input{src/fig/td_ss}

To advance CPS research, we need ways to systematically study CPS when varying factors, allowing comparison of CPS results across settings.  
We therefore develop a CPS task environment generator, \cpsrp, which can generate CPS environments with different design factors.
We also release a CPS task design checklist, \cpscheck, that describes how varying design factors produces different environments.  We defer discussion of \cpscheck to Section \ref{sec:cps-check}; here we give a concrete description of the task environments our work targets.

We start with several requirements:
(R1) \cpsrp{} should be built on an activity that can support the different values in \cpscheck; 
(R2) the activity should be \textbf{fun}, to motivate participant signups, because CPS studies require multiple participants, making scheduling 
a logistical barrier to conducting CPS research;
 (R3) the activity should be easy to learn for both participants and researchers, in order to minimize time spent in tutorials and allow researchers to quickly design different CPS studies;
 and (R4) the activity should easily scale in difficulty to enable CPS research studying effects of expertise on collaboration.

We meet our design requirements by using the Tower Defense (TD) game genre as our \cpsrp{} activity. The premise of a TD game is to defend a base from enemies by placing towers on the map, which can destroy the enemies. TD games require strategy and resource management---a vital aspect of CPS tasks~\cite{care2015collaborative}---and games have been successfully used by the research community to study communication (e.g., Codenames;~\cite{shaikh2023modeling}) and collect data (e.g., Verbosity; \cite{verbosity}, Duolingo~\citep{duolingo}, SearchWar \citep{searchwar}, and MatchIn~\citep{matchin}. 

TD games are known for having a gentle learning curve, short levels (R3), and ease in scaling difficulty through simple designs (R1, R4;~\citealp{avery2011computational}). The 2021 mobile market value  for TD games was estimated at 940 million USD~\cite{mobile-analytics}; this popularity suggests the potential for participants to play the game of their own volition (R2).
It is also known to support 1--4 players in cooperative play,\footnote{Bloons TD 6~\texttrademark{} is a commercial game with a 4-player cooperative mode.} natively supporting studying human-AI teams involving as few as one human.

We briefly describe what a TD game involves, referencing an in-game screenshot (\autoref{fig:td_ss}) of an environment produced by \cpsrp.
In a TD game, the player needs to defend their base (7) from enemies by placing towers on the map whose inhabitants can attack the oncoming enemies. The enemies will appear at designated spawn points (1) and traverse the map along specific paths known to the player, allowing the player to strategize where to place towers effectively.
Players must manage their resources (3) (e.g., gold and map real estate) when developing their defense strategy.
Levels differ in the enemy spawning behavior (e.g., enemies can spawn without a break, or there is time in between groups of enemies), enemy variants (e.g., a faster or slower enemy), map terrain (e.g., obstacles can prevent tower placement), and player resources (e.g., types of towers, amount of starting gold).
The standard TD game has two phases: \emph{planning}, a static phase where players can place towers on the map, and \emph{attack}, a dynamic phase during which enemies spawn, and players can react to the changing situation by adjusting their towers.

\cpsrp{} is built on the open-source Godot\footnote{\url{https://godotengine.org}} game engine, and further details of implementation and the tower defense games it produces are available in ~\autoref{app-system-overview} and the documentation of our open-source release.

%% file: src/fig/td_ss.tex
\begin{figure*}[!th]\center
    \includegraphics[width=0.9\linewidth]{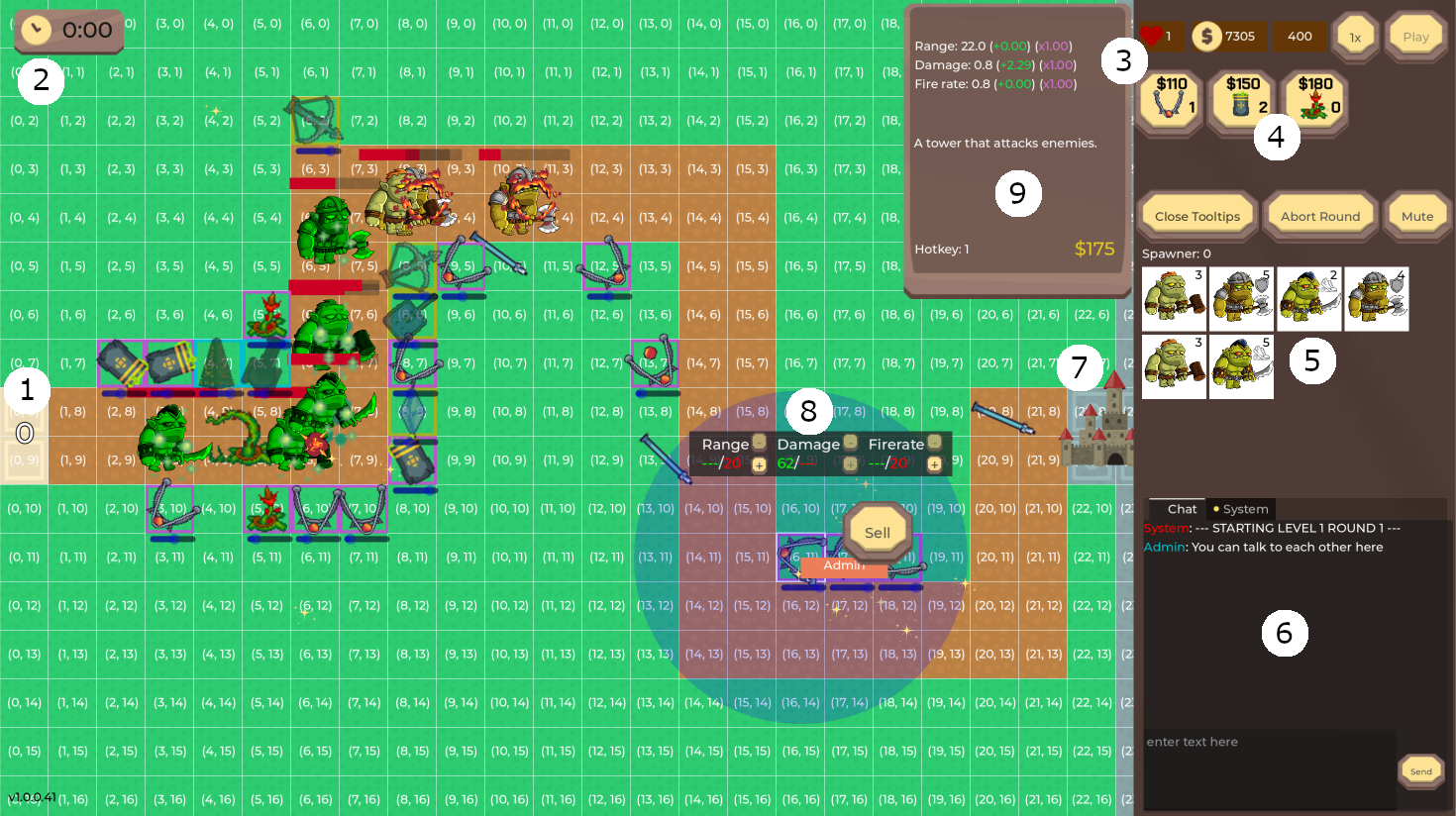}
    \caption{In-game screenshot of a game produced by \cpsrp, used in our case study. Enemies spawn from (1) and can only move on the brown path. Towers can only be placed on the green spaces. (2) is the timer used during the \emph{planning} phase, indicating how much time players have to set the board before the \emph{attack} phase starts. (3) tracks base health---players lose if it drops to zero due to enemies reaching the base, the amount of money available to purchase towers and upgrades, and a running score. (4) is the set of towers this player can build. Different towers have different abilities and costs. (5) previews the enemy sequence of a spawn point. (6) is the text chat players use to communicate with each other. (7) is the base players must defend. (8) is an upgrade menu for a selected tower. (9) is an information panel about a tower. A coordinate grid is provided so players can refer to specific spaces on the map when communicating with each other.}
    \label{fig:td_ss}
\end{figure*}

%% file: src/10_cps_checklist.tex
\input{src/tab/design_qs}
\section{\cpscheck~: A CPS Task Design Checklist}\label{sec:cps-check}

The PISA 2015 CPS Framework (\pisa) \citep{pisa-2015} describes CPS tasks through a set of 15 design factors, showing how different CPS settings can be studied by manipulating different combinations of factors (e.g., team size and composition). 
To operationalize CPS research goals as design parameters that \cpsrp{} can use to generate the environment, we define \cpscheck, a design checklist adapted from \pisa{} (\autoref{tab:design_qs}). We provide default values for \cpscheck{} items in the event that some items are unnecessary to adjust for a particular study.
We next explore how different hypothetical  research goals can be targeted with different TD games generated by \cpsrp{} and designed with the help of completing \cpscheck{}.

\paragraph{Goal: Compare solution quality between all-human teams and mixed human-AI teams.}
To compare solution quality, we require a more complex task evaluation function than a simple binary win/lose value (Q1). We can design a scoring function to incorporate the time required to agree on a strategy during the planning phase, the amount of money used, or the distance enemies travel. We can also adjust the solution space size (Q7). A level can have a single solution, requiring a specific strategy for placing towers, and solution quality is evaluated by the speed of figuring out the solution. A level can also have multiple solutions, with solutions rated for quality, e.g., a solution using the minimum amount of towers is harder to achieve than a solution maximizing resource consumption and is thus higher quality. The solution quality comparison between teams can then measure the rate of solving levels with minimal resource consumption.

We want to use team compositions with different fractions of human and AI players (Q4). 
We can investigate how different team roles and personalities in all-human or mixed human-AI teams affect solution quality (Q5); for example, an all-human team where everyone identifies as a leader and has the same towers could result in poor solution quality due to an increase in conflict over strategy; or a team where a human leader effectively uses support towers from an AI teammate (Q6) may outperform a team with an AI leader who does not request support towers from a human teammate.  Since we are interested in manipulating team composition, we can give all players a shared resource pool so that information is updated and distributed to all players simultaneously (Q8).

\paragraph{Goal: Investigate how stress affects team performance and communication.} 
Stress can affect team performance, learning, and communication~\citep{pfaff2012negative,savelsbergh2012team,orasanu2004team}, with more successful teams developing adaptive strategies \cite{kontogiannis1999stress}. 
We can model stressful situations by adjusting the amount of starting resources (money and planning time) to require more dynamic gameplay during the attack phase, forcing players to adapt to a rapidly changing environment (Q9). To design levels requiring more dynamic gameplay, we limit the initial starting resources such that players cannot beat a level by only placing towers during the planning phase. As enemies are defeated, players gain additional gold to spend towards placing more towers and upgrading existing towers, which are required to successfully defend their base. 
The control condition can then be giving players plentiful starting resources.
We will evaluate the task with a simple binary win/lose (Q1) and allow several possible solutions so that teams are not discouraged if they cannot land on the single most optimal solution (Q7).
Giving less money and planning time means players have to monitor the changing situation during the attack phase.
We enable voice communication (Q11) so that typing speed is not a factor.

\paragraph{Goal: Reimplement and extend prior work.} \label{sec:reimp}

Although \cpsrp{} is designed to generate TD games, we can simulate object selection and manipulation tasks by limiting player interaction.

\textbf{Object Selection}. Reference games used in KTHTangrams~\citep{shore-etal-2018-kth} and PentoRef-Take~\citep{zarriess-etal-2016-pentoref} are played with two players in the roles Instruction Giver (IG) and Instruction Follower (IF). Both players have a view of the map. The IG is given the game goal (select a specific piece), and the IF can manipulate the map (select the piece). 
We simulate this task using \cpsrp, by designing levels with towers placed on the board at the start, replacing the tower imagery with a pentomino or tangram.
We enable voice communication and end the level upon a single tower object selection, evaluating success through whether the correct tower was selected (Q1).

\textbf{Object Manipulation.} \citet{dolldialogue} designed a task for furnishing a physical dollhouse. The IG is given the furnished dollhouse, and the IF is given an empty house. The IG needs to instruct the IF to furnish the house, and task success is evaluated by the correctness of object location and orientation. 
To simulate this task in \cpsrp, we design levels that resemble house interiors, with walls designating rooms and preventing towers from being placed on them. We give the IF a set of 
towers that can be placed in the level, replacing the tower imagery with furniture. A tower can span multiple grid spaces on the map, and there are multiple copies of each tower with different orientations. The IG is provided the same level but with towers placed on the map already (similar to the setup for the reference games). 
Voice chat is enabled for communication.
Since \cpsrp{} produces digital grid-based games, object location and orientation can be automatically evaluated for correctness, improving upon the original setting, where evaluation was manually coded. 
A limitation of our simulation is that the original task used a physical dollhouse, giving participants multiple perspectives of the board (which could increase task complexity), while our simulation only gives players a single top-down view. 3D simulations or creating multiple 2D perspectives could be explored in future work.

%% file: src/tab/design_qs.tex
\begin{table*}[!thb]\centering
	\scriptsize

	\begin{tabular*}{\linewidth}{@{\extracolsep{\fill}} p{0.2\linewidth}  p{0.4\linewidth} p{0.4\linewidth}}\toprule
		\multicolumn{3}{c}{What are we studying? E.g., Decision making, collaborative learning, negotiation, exploratory group work, how stress affects communication} \\ \midrule
		Context                               & Dimension                                                                                                                                     & Example Values                                                                                                                                                \\ \midrule
		        
		\multirow{4}{*}{Problem Scenario}     & \multicolumn{1}{l}{Q1. How is the task evaluated for success?}                                                                                & Binary win/lose, score(time, health)                                                                                                                          \\
		                                      & \multicolumn{1}{l}{\textcolor{red}{$^\star$}Q2. How long does one CPS instance take to complete?}                                             & 1 minute for planning and 1 minute for attack                                                                                                                 \\
		                                      & \multicolumn{1}{l}{\textcolor{red}{$^\star$}Q3. How do skill and expertise scale with repetition?}                                            & \begin{minipage}[t]{0.7\linewidth} Levels of similar difficulty are repeated, level difficulty scales by introducing more enemy spawn points \end{minipage} \\\midrule
		\multirow{3}{*}{Team composition}     & \textcolor{red}{$^\star$}Q4. What fraction of teammates are human or AI?                                                                      & H-H-H, H-AI, H-AI-AI, H-H-AI                                                                                                                                  \\
		                                      & Q5. What is the symmetry of roles?                                                                                                            & \begin{minipage}[t]{0.7\linewidth}2 players have the same support towers, and 1 has all offense towers    \end{minipage}                                                                                      \\
		                                      & Q6. How are teammates interdependent?                                                                                                         & Support towers are necessary to beat the level                                                                                                                \\\midrule
		\multirow{4}{*}{Task characteristics} & Q7. How open is the solution space?                                                                                                           & Only 1 tower placement configuration can win                                                                                                                  \\
		                                      & \begin{minipage}[t]{0.7\linewidth} Q8. What information is available, and how is new information distributed (if applicable)?\end{minipage} & \begin{minipage}[t]{0.8\linewidth}All players have the same information at all times, players must discover enemy spawn sequence\end{minipage}              \\
		                                      & Q9. How much stress are players under?                                                                                                        & No stress (unlimited planning time)                                                                                                                           \\
		\midrule
		Medium                                & Q10. What is the communication medium?                                                                                                        & Text, voice                                                                                                                                                   \\
		\bottomrule
	\end{tabular*}
		\caption{\cpscheck: Design questions adapted from PISA 2015 CPS design contexts. 
		Questions with \textcolor{red}{$^\star$} are added to help design studies where task repetition is a dependent variable 
		or considerations for human-AI teams. 
		H = human.
	}
	\label{tab:design_qs}
\end{table*}

%% file: src/7_case_study.tex
\section{Case Study: Communication of 
Small Groups as Task Difficulty Increases}

To validate its flexibility, we want to explore whether \cpsrp{} is capable of producing an environment that elicits diverse collaborative problem solving behavior.
Prior work in CPS primarily used tasks with dyads or task reptitions at the same difficulty level, so we design a CPS task where teams of 3--4 people complete a task, aiming to minimize expenditure of gold, at multiple difficulty levels.

We design our \cpsrp{} environment as follows, referencing the questions from \cpscheck.
Task success is evaluated by the amount of money left unused, enemies destroyed, and health of the base (Q1).
A single level takes 5--8 minutes to complete, depending on level difficulty, and we design 3 levels with increasing difficulty (Appendix~\autoref{fig:td_levels}; Q2--3).
All players are human (Q4), and each player is given 2--4 unique towers from a pool of 12 towers with different properties (\autoref{app:td_params}) so that players have different roles, encouraging all players to engage and suggest usage of their own towers (Q5--6).
Players are provided a surplus of gold, and costs are balanced to slightly favor upgrading over placing more towers, giving teams the opportunity to find many successful strategies (Q7).
All new information is distributed to players simultaneously (e.g., how much damage an enemy receives from a tower) (Q8).
Players are under moderate time stress because each level is calibrated to give ample but limited time (5--6 minutes) to discuss strategy and place towers, and we disabled interaction during the attack phase (Q9). Players could end the planning phase early. We designated level-specific planning time to ensure the study is completed in a reasonable amount of time.
Players can only communicate through text chat (Q10).
These design decisions showcase the simplicity with which the TD genre affords the ability to create different CPS task environments.

\subsection{Data Collection}
12 teams of 3--4 people (total 42 individuals) were recruited to participate in a 1.5-hour study\footnote{Our local IRB approved our study.} and compensated with a gift card at a rate of 20 USD/hour. 
The study was conducted both in-person and remotely, and all studies were moderated. Recruitment occurred through school email listings and paper flyers posted around town. Participants were aged 18--24 (72\%), 25-31 (18\%), and 32+ (10\%); 55\% of participants were current undergraduates and 36\% were in a graduate degree program; a third of participants rated their tower defense game familiarity below 3 on a 5-point Likert scale.
Familiarity between teammates was not controlled, allowing some team compositions to contain strangers and others a subset of friends.

The study began with individual pre-surveys collecting basic demographic information, then participants watched a tutorial video explaining how to play the game and played a simple tutorial level together to become familiar with the interface. After the tutorial, they were given time to ask any questions about how to play the game. They then played 3 different levels 3 times each for a total of 9 games. Subsequent levels increased in difficulty, but the three rounds were the same for each level. Finally, they completed individual post-surveys containing questions about teamwork quality, team role identity, and team communication.

%% file: src/4_analysis.tex
\input{src/tab/cps_skills}
We logged data using XML tags, and the data logged was text communication, score, and tower interaction (upgrading, placing, and selling). The metadata associated with the data was the coordinates of interacted towers, timestamps, and the user.
The first 4 teams were used to calibrate game difficulty and level designs 
and the data from one team was excluded from analysis because a team member left early, resulting in a final dataset of 7 teams producing 1.5k utterances with a vocabulary size of 1.2k (Appendix \autoref{tab:casestudy_stats}).

\subsection{Observations}
We adapt a CPS skill taxonomy developed by \citet{cps-math} to describe the communication data, simplifying the initial 10 skill taxonomy to 8 because of low annotation reliability (\autoref{tab:cps-skills}).\footnote{We discuss annotation challenges in Appendix Subsection \ref{sec:annotation-challenge}.} We label only explicit natural language communications---the original taxonomy also includes system interactions (e.g., the act of placing a tower could be classified as ``executing action'').
A sample of 45 utterances of the data was manually annotated by two authors (inter-annotator agreement of 73\%), then one author annotated 3 games (30\% of the data). 
Example team communication is in Appendix \autoref{tab:cps-examples}, exemplifying planning and directing through natural language, as well as communication through game behavior (e.g., placing a tower at a specified location when requested without using language to acknowledge the request.)

Cognitive CPS skills were used 49\% of the time, and 29\% of all communication was devoted to developing strategic plans (planning and negotiation skills). \citet{cps-math} observed 30\% cognitive skill usage using a traditional collaborative math task, suggesting that the TD task in \cpsrp{} is a viable task for CPS studies.

From the surveys, we saw that the game was positively received, supporting our objective of developing a \emph{fun} CPS activity (R2). 43\% players commented that the game was fun, three players requested an official game release to play with others, and no player complained about task tedium.

\subsection{Analysis}
\input{src/fig/game_behavior}
Our levels were designed to give players a wide solution space through having an abundance of gold (e.g., Level 1 could be completed with 14k gold unspent).  This design emphasized problem space exploration over negotiating for a single optimal solution and is reflected in the low ``negotiation'' skill usage (4\%) and high  spread of placed towers (Appendix \autoref{fig:td-placement}). 
\autoref{fig:game-behavior} shows an example of two teams solving Level 2 with different strategies in tower placement and quantity. One team chose to concentrate their towers where the two paths meet so that towers can attack enemies on both routes, while another team placed many towers across the whole map. Our scoring function emphasized minimizing expenditure, so \autoref{fig:team-a} received a higher score than \autoref{fig:team-b}. Rounds were repeated three times, allowing teams to optimize working solutions--however, teams did not learn to significantly change expenditure behavior, which suggests cautious game behavior (Appendix \autoref{fig:money-unspent}). Teams 1 and 5 appeared to be confused about the task goal, often spending more money across rounds despite winning a previous round.

%% file: src/tab/cps_skills.tex
\begin{table*}
	[!tp]\centering
	\scriptsize

	\begin{tabular}{rrrrrr}\toprule
		Dimension                  & CPS Skill                                             & Example                                             	&	Count 	& Avg. Tokens \\ \midrule
		\multirow{4}{*}{Social}    & \multicolumn{1}{l}{Maintaining communication}         & ``haha okay''                                       	&	222		&	2.3 \\
		                           & \multicolumn{1}{l}{Sharing information}               & ``I have a tower damange all enemies''               	&	114		&	7.0\\
		                           & \multicolumn{1}{l}{Establishing shared understanding} & ``what does the diamond tower do?''                  	&	67		&	5.4\\
		                           & \multicolumn{1}{l}{Negotiating}                       & ``do we want to risk getting rid of anything else?'' 	&	38		&	5.4\\ \midrule
		\multirow{3}{*}{Cognitive}                           & \multicolumn{1}{l}{Representing and formulating}      & ``fires in multiple directions'' 	&	105		&	9.3                    \\
		                           & \multicolumn{1}{l}{Planning}                          & ``ok we can chokepoint the corners''                	&	227		&	7.2 \\
		                           & \multicolumn{1}{l}{Executing Actions}                 & ``k i maxed [upgrades]''                             	&	42		&	5.9\\
		                           & \multicolumn{1}{l}{Monitoring}                        & ``50 seconds D:''            	&	86		&	5.0	 \\
		        
		\bottomrule
	\end{tabular}
 	\caption{CPS Skill usage from our case study. Descriptive statistics are from the human annotated data (30\% of the full dataset). Utterances were tokenized using the Spacy en\_core\_web\_sm model.
	}
	\label{tab:cps-skills}
\end{table*}

%% file: src/fig/game_behavior.tex
\begin{figure}[!t]
    \centering
    \begin{subfigure}[b]{0.45\textwidth}
        \centering
        \includegraphics[width=\textwidth]{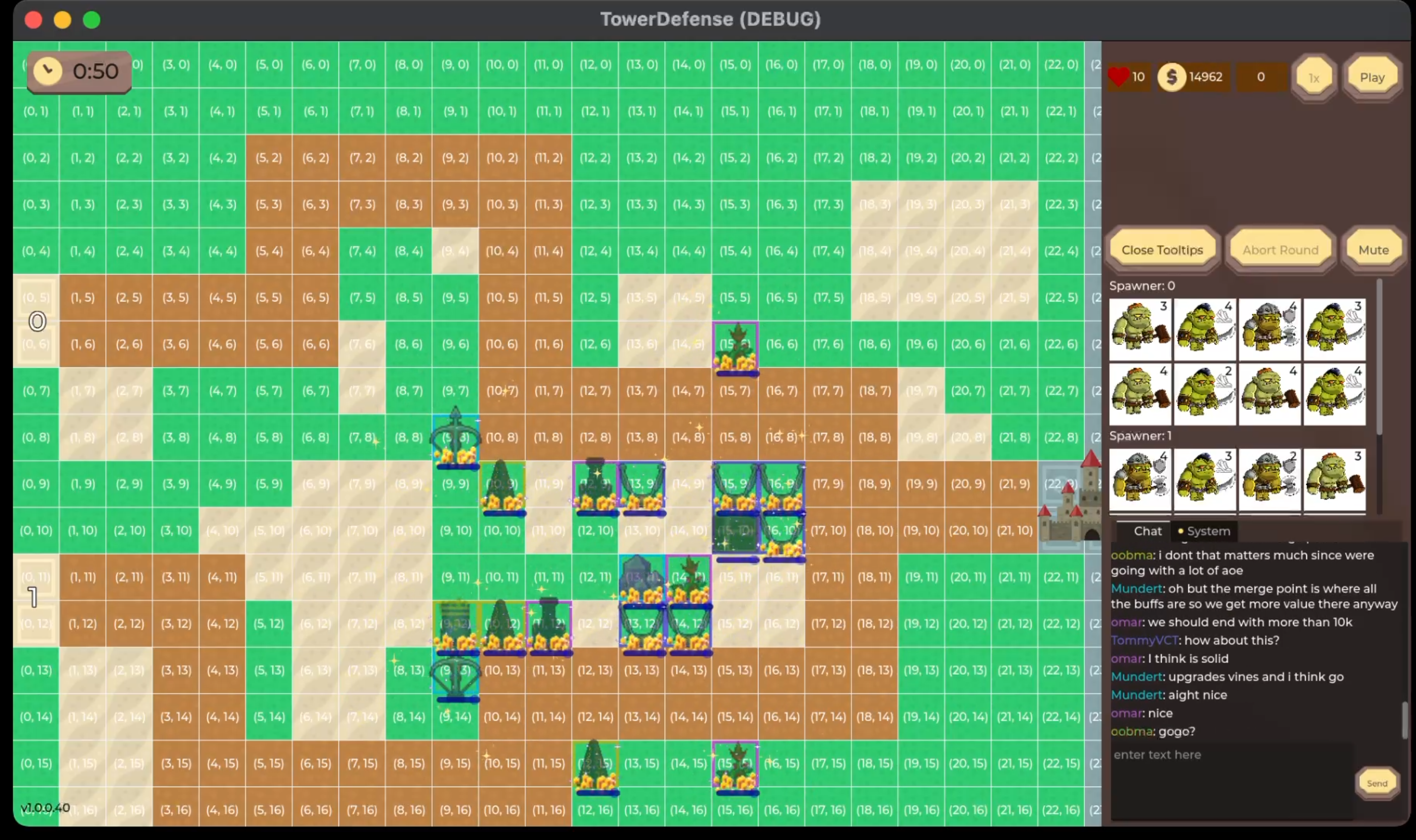}
        \caption{}
        \label{fig:team-a}
    \end{subfigure}
    \hfill
    \begin{subfigure}[b]{0.45\textwidth}
        \centering
        \includegraphics[width=\textwidth]{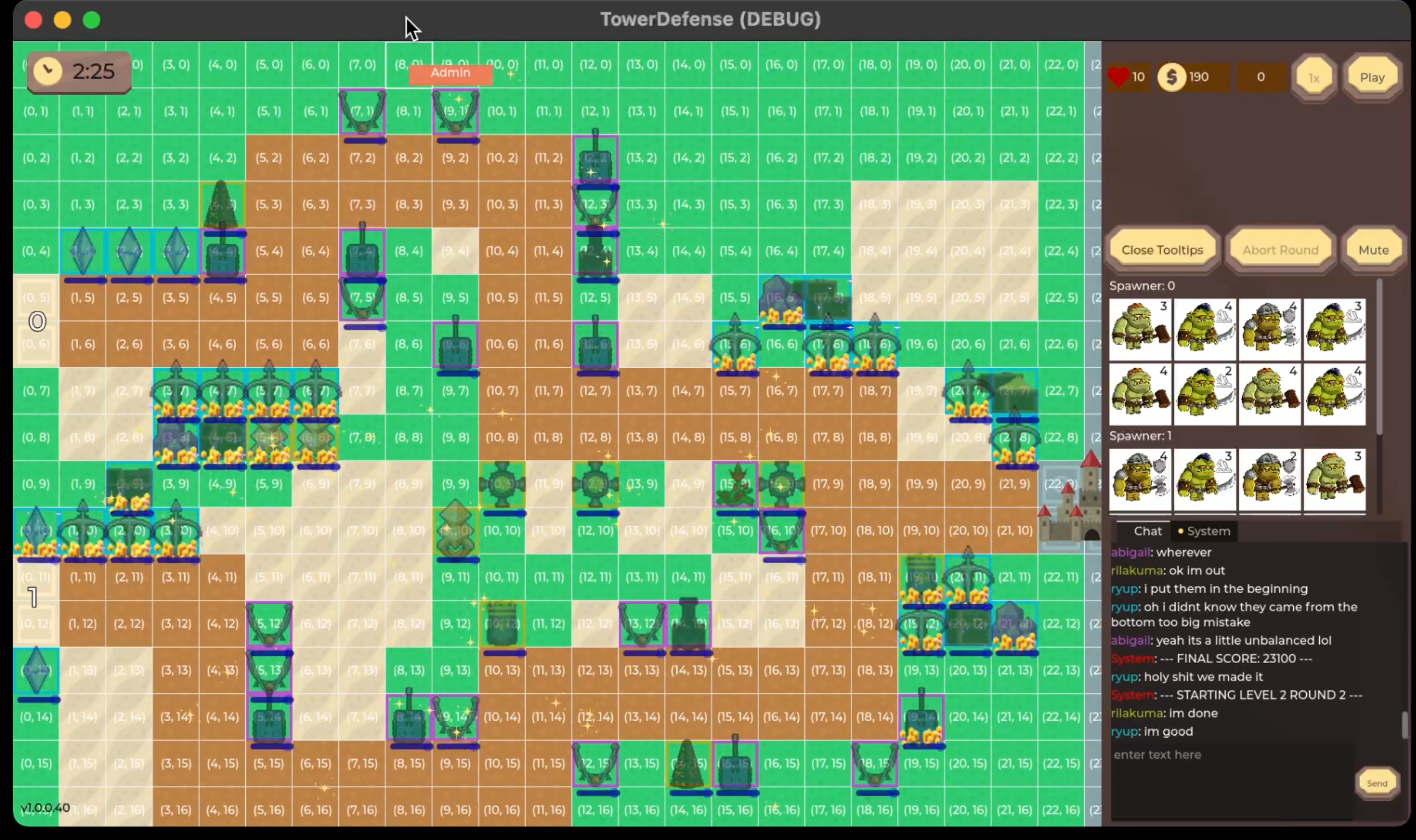}
        \caption{}
        \label{fig:team-b}
    \end{subfigure}
    \caption{Different strategies that succeeded in level 2. Players in (a) spent less and placed fewer towers. They concentrated their towers where the two paths converged, while players in (b) used the full map.}
    \label{fig:game-behavior}
\end{figure}

%% file: src/11_related_work.tex
\section{Related Work}
Prior work in CPS has studied a range of factors to understand effective teams, from identifying the effects of team member personalities on team outcomes to how teamwork processes can be evaluated.
When an AI teammate is involved, an important research direction investigates how and why humans choose to rely on AI. Findings from CPS human team processes can lead to improvements in AI agents and discovering how to better integrate AI into human teams to solve more complex problems.

Researchers have investigated how team composition affects human team outcomes (e.g., \citealp{ruch2018team,mathieu2014review,bell2018team,hollenbeck2004bridging}, \interalia), discovering predictors of team outcomes through team roles, individual expertise, demographics, and team knowledge.  
\citet{personality-matters} found five-person teams  with balanced personalities outperformed those with an imbalance in personalities on collaborative tasks. 
Analogously, \citet{wang2023unleashing} and \citet{fan2024ai} were able to improve  LM performance on downstream tasks by instructing the LM to simulate teams of domain-specific personas to collaborate internally. 
Priming an LM agent with a persona enables the simulation of inherited knowledge and linguistic patterns~\cite{Masumura2018RolePD,Wei2023MultiPartyCC,Park2023GenerativeAgents}, and searching for optimal personas in \hai{} teams could lead to improvements in  \hai{} team performance.

CPS tasks can be evaluated for overall task success, but improving teamwork requires evaluating intermediate processes. 
\citet{PAVEZ2022951} analyzed over a hundred studies on team performance measurement to propose a framework for evaluating teamwork along 4 dimensions: project team processes, project team emergent states, project team tangible outcomes, and project team perceptual benefits.
Educators have classified CPS communication for CPS skill usage to provide feedback to students on how to improve their group communication~\citep{cps-math,circuit,flor-etal-2016-automated,CPSCoach}.
Despite extensive work in evaluating CPS teams, there is little data released to the research community.

Research in AI-assisted decision making has produced valuable insights into how humans rely on AI advice. 
AI is increasingly involved in high-stakes decision, e.g., medical diagnoses, which has led to work in trust and reliability of AI. 
Humans are known to overrely on AI, following AI suggestions even when they are wrong~\citep{10.1145/3287560.3287590,671659,7349687}.
As a result, designing methods to encourage appropriate reliance on AI advice is vital, such as studying the effects of AI explanations \cite{10.1145/3640543.3645210,fleis-algo-aversion,10.1145/3411764.3445717,vasconcelos2023explanations}.
\citet{GAZIT2023100009}, \citet{Mesbah2021WhoseAC}, and \citet{10.1145/3653708} designed studies to understand human (over)reliance on AI using “judge-advisor system” (JAS) tasks where a human or AI advisor provides advice to a human judge, and the judge is responsible for making the final decision. 
However, decisions in these tasks are independent, and the judges are not able to explain their reasoning to the advisor in a bid to adjust the advisor's position, preventing the study of longer-term effects of \hai{} interactions and \hai{} communication.
Furthermore, the JAS task setup is traditionally dyadic, with one human and one (AI) advisor. In an exploration of {\em group} decision making, \citet{10.1145/3640543.3645199} recruited groups of two people to follow the judge-advisor system with an AI advisor. They then introduced an AI agent to play devil's advocate and found the agent successfully encouraged more appropriate reliance of AI advice.

%% file: src/6_discussion.tex
\section{Conclusion}
Human-AI collaborative problem solving tools are rapidly being integrated in real-world work environments. The modern workforce uses teams with more than two parties, but empirical research with larger teams lags behind. The task design space for conducting CPS research is large, and the tooling to systematically explore CPS designs is lacking. Our CPS task environment generator, \cpsrp, enables diverse, systematic CPS research through a tower defense game environment that appeals to human subjects and is grounded in theory. It enables the study of larger team CPS (multiple people and/or multiple AI agents) grounded in an environment and task that is accessible yet still carries real-world resemblance.
The data generated in our case study reveals different collaborative tasks required to succeed in the overall tower defense task, such as decision making and ensuring teammates have the same understanding of the task. 

We will release all code for \cpsrp{} and communication data collected in our case study to encourage studying multi-human and multi-AI collaborative problem solving.

%% file: src/z_extras.tex
\section{Limitations}
The tower defense task in \cpsrp{} environments has a learning curve (albeit a gentle one), so tutorials and practice before the actual study commences may be longer than simpler tasks such as a reference game. This complexity is necessary to support a broad range of complex tasks. \cpsrp{} environments currently only support a top-down perspective of the world, so supporting first-person settings (e.g., simulating a Minecraft search and rescue task) is infeasible. We believe these design limitations can encourage the development of other similarly specialized CPS environment generators.

Our initial release of \cpsrp{} implements many common attributes of tower defense games. There are many more attributes available for implemention that have been successfully deployed in commercial tower defense games that may be beneficial for future CPS studies, such as increasing the task difficulty by giving enemies resistance to certain towers. We hope to see \cpsrp{} evolve in its feature set through usage.

Although \cpsrp{} was developed in English, and our case study used English, usage of \cpsrp{} does not require English. Our case study also required using text communication, however \cpsrp{} does not limit the study of CPS to text communication settings. \cpsrp{} was built in the open-source game engine Godot which natively supports other languages, localization, and microphone input. At this time, expanding to video and other modality inputs is not supported.

\cpscheck is adapted from \pisa{}, but the CPS researcher may find other CPS frameworks (e.g., ATSC21, \citealp{hesse2015framework}, and the generalized competancy model by \citealp{generalized-cps}) more appropriate as a checklist. We expect adapting other frameworks into a checklist that can be used to generate \cpsrp{} environments should not be a major challenge, as other frameworks are describing CPS tasks using different attributes, and the TD game used in \cpsrp{} is fundamentally a CPS task.

\section{Ethical Considerations}
The flexibility in designing CPS task environments through \cpsrp{} necessarily places a large responsibility on the designer to design studies appropriate for their target audience or research goal. For example, the imagery used in-game for enemies and towers could be offensive to certain audiences and should be adapted as needed. As with any study in communication, appropriate content filter measures should be in place as required.

The development of generative AI agents as peers that can communicate with humans comes with the risks of the AI agents generating inappropriate content and the concerns of AI replacing humans. Our intentions are that the AI agents can augment human capabilities in more complex problem solving situations, boosting CPS abilities; however, we acknowledge that some problem solving tasks can be simulated and solved through internal or multi-agent collaboration.

Our study was approved by our institution's IRB, and participants were fairly compensated and consented to data sharing with the research community.

%% file: src/x_appendix.tex
\section{\cpsrp{} System Overview} \label{app-system-overview}
\cpsrp{} is built using the open-source game engine Godot,\footnote{\url{https://godotengine.org}} Nakama,\footnote{\url{https://heroiclabs.com/nakama/}} and data collection uses REST API calls to an external server.\footnote{The external server we release alongside \cpsrp{} is a Python Flask server.} All code within Godot is written in GDScript. 
Godot has native support for multiplayer networking, text localization, and game design content can be saved to human-readable text-based formats, allowing researchers to design environments with minimal knowledge of Godot. 
It also has an active plugin ecosystem that enables easy extensibility, including AI agent plugins (e.g., Godot RL Agents~\cite{beeching2021godotrlagents} and GodotAgent\footnote{\url{https://github.com/Wizzerrd/GodotAgent}}) for conducting human-AI research.
Multiplayer syncing and logic is handled server-side, e.g., the server communicates the game state to clients, rather than game logic being computed on the client,and the client communicating to all other clients the updated game state.
For example, suppose a client player wants to upgrade a tower. The player interacts with the upgrade button, which sends a purchase request to the server. The server determines if the purchase is permissible, then communicates to all clients the new game state (an upgraded tower, if the purchase was permitted).  
Player game interactions (e.g., purchasing, upgrading, and selling a tower), communication, and game scores are logged to the external server by default. Additional data logging can be added as needed.
\cpsrp{} supports moderated sessions, where the researcher can enter the game to observe gameplay without acting as a player, and unmoderated play, where players can run sessions on their own. The game host is designated as the server for multiplayer, and a client player can simultaneously be the server.

\input{src/fig/td_system_overview}
\subsection{User Experience}
The experience flow is depicted in \autoref{fig:td_system_overview}, which we describe here. 
First, the game executable is distributed to all players.  Players authenticate through Nakama, then either a player or the experimenter (in a moderated session) hosts a game room. The host distributes the unique room key generated by Nakama to all other players. Players join the room and see a random team name that they can edit. The purpose of the team name is to improve team cohesion and collaboration through the construction of a group identity \cite{TeamBuildinginanExerciseSetting}. After all players have joined the room, the host starts the game. Players then play levels as designed by the experimenter (e.g., one level or multiple rounds per level).
At the end of a round, a leaderboard is displayed with the team name and score breakdown. Leaderboards are known to improve user performance \cite{mekler2013points,landers2017gamification}, and it allows teams to track their progress against themselves (for tasks with multiple rounds per level) and others.

\paragraph{User Interaction.} Each player is given a unique color that is used in the text chat display. The color is also used to outline the towers they placed (\autoref{fig:td_ss}; purple color) to indicate who placed which tower. 
Towers can be placed by clicking a button (\autoref{fig:td_ss}; 4) or through the assigned hotkey. Tower information is shown in a panel (\autoref{fig:td_ss}; 9) that appears when any tower is targeted. Selecting a tower will open an upgrade panel. Upgrades are given extra visual effects to help players understand the game state and mechanics~\cite{zhou2022datafeelexploringvisual}: upgrading the range that a tower can interact with alters the size of a colored circle around the tower, damage upgrades are indicated by the quantity of sparkles surrounding a tower, and firerate is shown through the speed of the orbiting sparkles. The addition of visual effects gives players an idea of which upgrades are applied to towers without needing to target towers to open the information panel.

\paragraph{CPS Interface Designs.} To facilitate CPS communication behavior, we include several user interface design parameters not commonly found in TD games that can be toggled and customized as needed.
Tower names can be hidden, which creates a setting similar to those used in common ground building studies, as players will need to develop a code to refer to specific towers.
We provide a preview of the sequence of oncoming enemies from a spawn point (\autoref{fig:td_ss}; 5), which is vital to experiments conducted without the dynamic attack phase. The preview gives information that players can use to plan their strategy, and enables longer level designs without requiring players to memorize the enemy spawn behavior if players can play a level multiple times. 
We provide a coordinate grid label across the map so that players can refer to specific locations, in a similar manner to chess coordinates. 
Features can be disabled depending on the experimenter's study goal, e.g., if the research goal is to investigate how different teams refer to a particular location, the experimenter may want to disable the coordinate grid label.

\subsection{Tower Defense Designs}\label{app:td_params}
Currently implemented tower defense designs that can be adjusted to suit the specified CPS task are as follows.

\begin{enumerate}
    \item Communication: Voice (bool), push-to-talk (bool), text chat (bool)
    \item Description visibility: Tower name (bool), tower description (bool)
    \item Number of rounds per level (int)
    \item Player resources: Money (shared or individual), health and Score (shared)
    \item Interactability during attack phase (bool). Enable this to allow adjusting tower placement and upgrading towers during the dynamic attack phase.
    \item Towers:  We provide 12 custom towers with unique mechanics and effects. Information about towers (name, description) can be customized. The unique towers are: basic, poison (damage over time), piercing (damage multiple enemies in a straight line), splash (area damage), obstacle (spawn an object on the track that does damage when enemies walk over it), slow (slows enemies), fear (enemies go backwards along the track), sniper (does more damage to faster enemies), discount (lowers upgrade costs of nearby towers), support (buffs all stats for nearby towers), multishot (shoots in 4 directions). 
    \item Levels: A level design designates how enemies spawn, the enemy movement paths, the location of a base that players defend, terrain for where towers can be placed, starting gold and health, and which towers are available to players. 
    \item Enemies: There are enemy variants that differ in health, movement speed, point value when destroyed, and money given to players when destroyed.
\end{enumerate}

We expect to implement other common game design paradigms such as segmenting the map so players can only place towers on their designated section as the platform matures.

\section{Case study results}\label{app:casestudy_params}
\input{src/fig/td_levels}
\autoref{tab:casestudy_stats} describes our case study in the context of other tasks with open data. \autoref{fig:game-behavior} depicts the levels and tower placing behavior in our case study. Sample conversations are in \autoref{tab:cps-examples}.
\input{src/tab/corpus_stats}

\clearpage
\input{src/tab/example_cps_data}
\clearpage
\input{src/fig/money_unspent}

\section{Survey Questions}
\label{app:surveys}
The pre-survey collected basic demographic information.
\input{src/fig/presurvey}
\clearpage
The post-survey contained the Teamwork Quality questionnaire~\citep{hoegl2001teamwork}, VIA Team roles inventory~\citep{ruch2018team}, and an open-ended task-specific questionnaire. Both TWQ and VIA used a 7-point Likert scale with options: Strongly Disagree, Disagree, Somewhat Disagree, Neutral, Somewhat Agree, Agree, and Strongly Agree.

\subsection{TWQ}

\begin{itemize}
    \item Communication
    \begin{itemize}
    \item There was frequent communication within the team
    \item The team members communicated mostly directly and personally with each other.
    \item There were mediators through whom much communication was conducted.
    \item Project-relevant information was shared openly by all team members
    \item Important information was kept away from other team members in certain situations.
    \item In our team there were conflicts regarding the openness of the information flow.
    \item The team members were happy with the timeliness in which they received information from other team members
    \item The team members were happy with the precision of the information received from other team members
    \item The team members were happy with the usefulness of the information received from other team members
    \end{itemize}

    \item Coordination
    \begin{itemize}
        \item The work done on subtasks within the project was closely harmonized.
        \item There were clear and fully comprehended goals for subtasks within our team.
        \item The goals for subtasks were accepted by all team members.
        \item There were conflicting interests in our team regarding subtasks/subgoals.
    \end{itemize}
    \item Mutual Support
    \begin{itemize}
        \item The team members helped and supported each other as best they could.
        \item If conflicts came up, they were easily and quickly resolved
        \item Discussions and controversies were conducted constructively.
        \item Suggestions and contributions of teammembers were respected
        \item Suggestions and contributions of team members were discussed and further developed.
        \item Our team was able to reach consensus regarding important issues.
    \end{itemize}
    \item Effectiveness
    \begin{itemize}
        \item Going by the results, this project can be regarded as successful.
        \item The team was satisfied with the project result.
    \end{itemize} 
    
\end{itemize}

Open-response questions:
\begin{itemize}
    \item What went well during the game?
    \item What went poorly during the game?
    \item Any notable communication difficulties or frustrations? If they were resolved, how did you resolve them?
    \item Any notable joyous or satisfactory communications?
    \item Suppose you played the game again with {\textbf different} maps but the {\emph same set} of players. What would you change?
    \item (Optional) Any other comments or complaints about your teamwork or communication?

\end{itemize}

\subsection{VIA Team roles}
Instructions for participants: for every role, read the description and answer the questions, imagining that you are currently in your {\textbf ideal} team.
\begin{itemize}
    \item Idea Creator. When working in a team, the creation of new ideas to come up with a solution for a difficult problem or task is essential. Thereby, Idea Creators are people with unconventional ways of coming to solutions and great ideas.
    \begin{itemize}
        \item In my ideal team, I'm at my best when coming up with ideas.
        \item I enjoy creating ideas within my ideal team
        \item I am able to be a great idea creator within my ideal team
        \item I have a feeling of energized focus when coming up with ideas within my ideal team
        \item It makes me feel good to create ideas in my ideal team
    \end{itemize}
    \item Information Gatherer. Information Gatherers search for information, for example on topics as best practices, new trends, potential vendors, competition, and so forth.
    \begin{itemize}
        \item In my ideal team, I'm at my best when gathering information
        \item I enjoy gathering information within my ideal team
        \item I am able to be a great information gatherer within my ideal team
        \item I have a feeling of energized focus when gathering information within my ideal team
        \item It makes me feel good to gather information within my ideal team
    \end{itemize}
    \item Decision Maker. Decision Makers are processing all the information at hand, integrating it to make the best possible decision and clarifying the goals.
    \begin{itemize}
        \item In my ideal team, I'm at my best when making decision
        \item I enjoy making decisions within my ideal team
        \item I am able to be a great decision maker within my ideal team
        \item I have a feeling of energized focus when making decisions within my ideal team
        \item It makes me feel good to make decisions within my ideal team
    \end{itemize}
    \item Implementer. Once a team has arrived at a decision on its direction, it needs to implement it. Thereby the Implementer constantly controls the current status and takes measures to work towards the goal.
    \begin{itemize}
        \item In my ideal team, I'm at my best when implementing goals
        \item I enjoy implementing goals within my ideal team
        \item I am able to be a great implementer in my ideal team
        \item I have a feeling of energized focus when implementing goals in my ideal team
        \item It makes me feel good to implement goals in my ideal team
    \end{itemize}
    \item Influencer. Commonly, the work product of the team needs to be presented by the Influencer for acceptance internally (supervisors, administrators) and/or externally (customers). This is a process of influencing and being persuasive.
    \begin{itemize}
        \item I'm at my best when representing the work/opinion of the team and convincing others of it
        \item As a member of my ideal team, I enjoy representing the work/opinion of the team and convincing others of it
        \item I am able to be a great influencer in my ideal team
        \item I have a feeling of energized focus when representing the work/opinion of my ideal team and when convincing others of it
        \item It makes me feel good to represent the work/opinion of my ideal team and convince others of it
    \end{itemize}
    \item Energizer. In the process of getting work done, Energizers are people that infuse energy into the work and others. Teams without enough energy can fall flat and struggle during times of pressure or prolonged projects that require endurance.
    \begin{itemize}
        \item In my ideal team, I'm at my best when energizing
        \item I enjoy energizing within my ideal team
        \item I am able to be a great energizer within my ideal team
        \item When I focus on infusing energy into work and others of my ideal team, I feel energized too
        \item It makes me feel good to energize within my ideal team
    \end{itemize}
    \item Relationship Manager. Since the working of a team is a dynamic interplay of people and their relationships, the Relationship Manager helps to run relationships smoothly and to resolve conflicts.
    \begin{itemize}
        \item In my ideal team, I'm at my best when managing relationships
        \item I enjoy managing relationships within my ideal team
        \item I am able to be a great relationship manager within my ideal team
        \item I have a feeling of energized focus when I manage relationships within my ideal team
        \item It makes me feel good to manage relationships within my ideal team
    \end{itemize}
\end{itemize}

\section{CPS classification}
The CPS skill taxonomy used for classifying utterances in the CPS pilot reproduced from ~\citet{cps-math}:
\begin{enumerate}
    \item Sharing information. Content relevant information communicated during collaboration and includes sharing one's own information, sharing task or resource information, and sharing understanding 
    \item Maintaining communication. Content irrelevant social communication and includes general off-topic communication, rapport-building communication, and inappropriate communication 
    \item Establishing shared understanding. Communication in the service of attempting to learn the perspective of others and trying to establish that what has been said is understood. 
    \item Negotiating. Communication used to express agreement or disagreement and to attempt to resolve conflicts when they arise 
    \item Exploring and understanding. Actions in the task environment to explore and understand the problem space.
    \item  Representing and formulating. Actions and communication used to build a coherent mental representation of the problem and formulate hypotheses 
    \item Planning. Communication used to develop a strategy or plan to solve the problem 
    \item Executing actions. Actions and communication used in the service of carrying out a plan (e.g., enacting a strategy or communicating to teammates actions one is taking to carry out the plan). 
    \item Monitoring. Actions and communication used to monitor progress toward the goal and monitor the team's organization
\end{enumerate}

\subsection{Annotation challenges}\label{sec:annotation-challenge}
Annotating the data for CPS skill using the taxonomy developed by ~\citet{cps-math} was challenging because labels did not have a clear distinction.

For example, consider the following snippet:
\begin{lstlisting}
(1) ManedWlf: I have a basic tower with a range of 22, fire rate of 0.8
(2) ManedWlf: Shall I place a couple close to the castle?
(3) tjwill:	Looks like we've got the same ones to start with, and sounds good!     
\end{lstlisting}

When ManedWlf describes the basic tower in (1), we can label the utterance for \emph{sharing information} because it is sharing resource information.
In (2), a plan is proposed to place some basic towers near the castle, which we can label for \emph{planning}.
In (3), we have an observation about both players having the same basic tower. This could be labeled for \emph{sharing information} because tjwill is sharing information about having access to the same basic tower. It could also be labeled \emph{representing and formulating} because tjwill is building a mental representation about how everyone has the same starting towers.

We defined a few soft rule for classification to help with annotation consistency, but we suggest future work should investigate designing a more complex taxonomy with clearer distinctions between labels. 

A few soft rules used when manually classifying CPS skills:
\begin{itemize}
    \item If a player asks for opinions about placing towers or making upgrades, classify it as Planning.
    \item If players agree to a plan, classify as Negotiating even if it's just ``ok'' because it is expressing agreement about a plan proposal.
    \item If a plan is proposed and another player proposes an alternative or disagrees, classify as Negotiation.
    \item Representing and formulating is about understanding the efficacy of towers or strategy enacted, e.g., ``the blue tower seems to slow enemies down''
    \item If a player asks someone else to do something, classify as Planning because it is working towards developing the strategy.
\end{itemize}

\subsection{Prompt}\label{app:gpt4-prompt}
We tried using automatic annotation with GPT-4, but annotation agreement was only 55\%, and developing a CPS classification model with higher accuracy is beyond the scope of this work. We list the prompt prefix used for documentation purposes. We used the prompt prefix to classify batches of 6 utterances.

\begin{lstlisting}[breaklines=true]
CPS skills list:
<skill>Sharing information</skill>. content relevant information communicated during collaboration and includes sharing one's own information, sharing task or resource information, and sharing understanding 
<skill>Maintaining communication</skill>. content irrelevant social communication and includes general off-topic communication, rapport-building communication, and inappropriate communication 
<skill>Establishing shared understanding</skill>. communication in the service of attempting to learn the perspective of others and trying to establish that what has been said is understood. 
<skill>Negotiating</skill>. communication used to express agreement or disagreement and to attempt to resolve conflicts when they arise 
<skill>Representing and formulating</skill>. actions and communication used to build a coherent mental representation of the problem and formulate hypotheses 
<skill>Planning</skill>. communication used to develop a strategy or plan to solve the problem 
<skill>Executing actions</skill>. actions and communication used in the service of carrying out a plan (e.g., enacting a strategy or communicating to teammates actions one is taking to carry out the plan). 
<skill>Monitoring</skill>. actions and communication used to monitor progress toward the goal and monitor the team's organization

You are given a numbered list of inputs. For each input:
Step 1: classify the <chat_text> for one or more <skills> displayed
Step 2: Explain your reasoning in <reason> tags.

Inputs
1. <speaker>ym2552</speaker> <chat_text>It's just when they come in big groups that's worrying, as it seems most towers can only focus on </chat_text>
2. <speaker<schou1</speaker> <chat_text>any chance we can get a buff or discount tower at 9,4?</chat_text>
3. <speaker>jane</speaker> <chat_text>willdo</chat_text>
4. <speaker>paul</speaker> <chat_text>hell, even 1 more turret near the bottom probably would've gotten them all, but we're doing good</chat_text>

Outputs
1. <skill>Representing and formulating</skill>
<reason>The speaker is explaining that when a lot of enemies come at once, they worry the towers will be overwhelmed.</reason>
2. <skill>Planning</skill>
<reason>The speaker is asking another player to place a buff or discount tower at a specific location to further develop the solution</reason>
3. <skill>Executing actions</skill>
<reason>the player is acknowledging a request to act, showing they will execute an action</reason>
4. <skill>Representing and formulating</skill><skill>Maintaining communication</skill>
<reason>the player hypothesizes having one more turret near the bottom would have helped the strategy, then comments the team is doing well to build rapport.</reason>
---
Inputs
    
\end{lstlisting}

\section{Potential \cpsrp{} Tasks}
We decided to use the tower defense game genre as the task for \cpsrp{} after considering several other games.

\begin{enumerate}
    \item Pandemic~\texttrademark{} board game. We found valuable play by forum games that demonstrated the type of multi-turn collaborative communication we hope to see in CPS data. However, one instance of the game takes at minimum 30 minutes to complete, making it challenging to evaluate intermediate task process. The lengthy duration is also a barrier to task repetition within a single study session.
    \item Cryptic Crossword puzzles. The cryptic crossword puzzle variant relies on metahints and wordplay, making it more accessible than regular crosswords that require trivia knowledge. However, learning the rules is difficult. Participants required 2--3 hours to understand the rules in pilot tests. The communication during the task was also often short utterances suggesting the solution, with reasoning provided only if teammates requested. 
\end{enumerate}

\section{License}
The Godot game engine has an MIT license.
The terms for use of our artifacts
will be included in our released package.

%% file: src/fig/td_system_overview.tex
\begin{figure*}[!t]\center
    \includegraphics[width=0.9\textwidth]{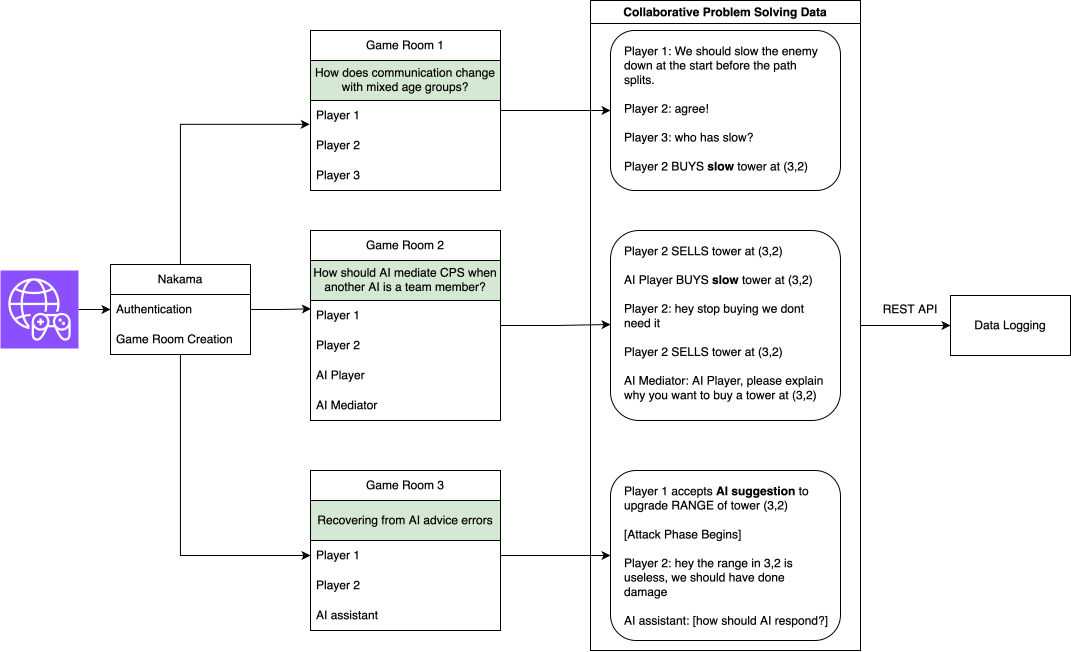}
    \caption{System overview illustrating 3 different research questions that \cpsrp{} supports. Players authenticate through Nakama, join game sessions with different experimental environment designs driven by research questions, and generate CPS data while playing the game. Player interactions and communication are collected using REST APIs.}
    \label{fig:td_system_overview}
\end{figure*}

%% file: src/fig/td_levels.tex
\begin{figure*}[!tb]
\centering
    \begin{subfigure}[b]{.9\textwidth}\centering
        \includegraphics[width=0.9\textwidth]{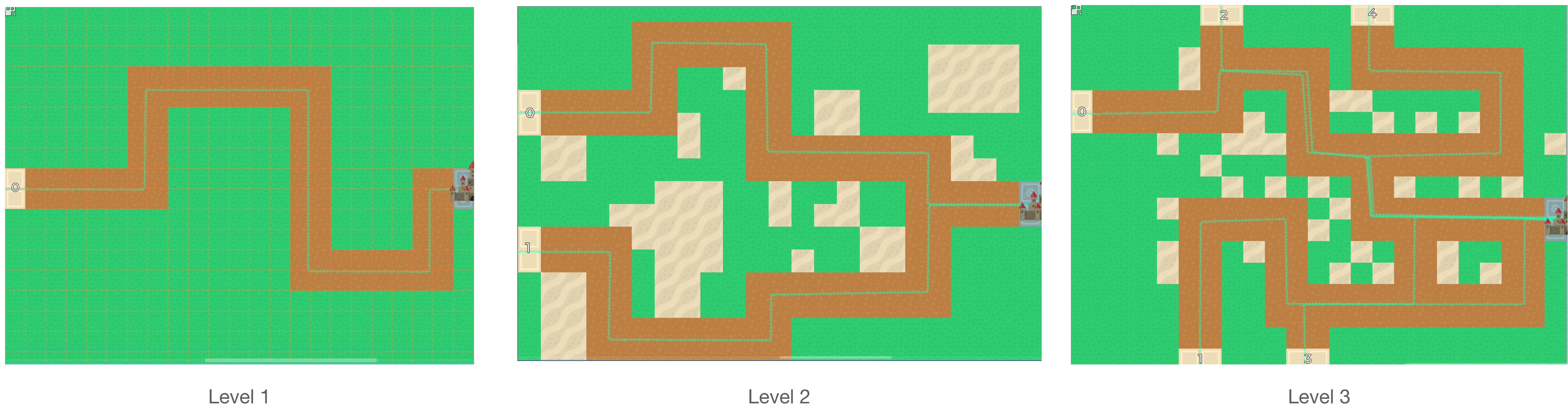}
    \caption{Level maps used in the \cpsrp{} case study. Players can only place towers on the green spaces. Enemies spawn at labeled spawn points and move along the brown paths to the castle on the right. Level difficulty was scaled by introducing more enemy spawn points and limiting the green spaces for tower placement. }
    \label{fig:td_levels}
    \end{subfigure}
    \begin{subfigure}[b]{0.9\textwidth}\centering
        \includegraphics[width=0.9\textwidth]{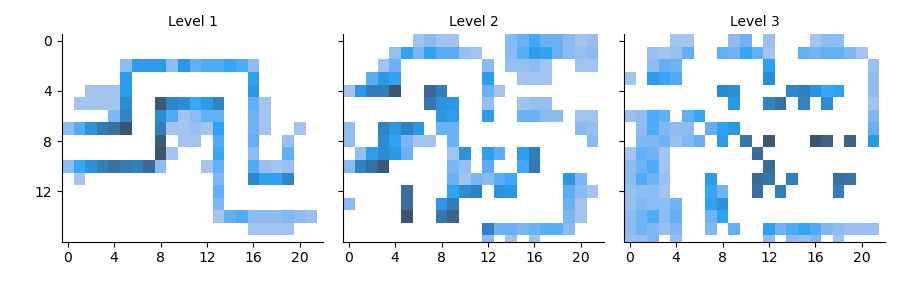}
    \caption{Tower placement frequency. Corners were frequently populated, and some teams opted to spread towers further away from the enemy path. Darker indicates higher frequency.}
    \label{fig:td-placement}
    \end{subfigure}
    \caption{Game levels and tower deployment in the \cpsrp{} case study.}
\end{figure*}

%% file: src/tab/corpus_stats.tex
\begin{table*}[!tp]\centering
	\scriptsize\setlength{\tabcolsep}{4pt}

	\begin{tabular}{lrrrrrrrrrr}\toprule
		& Teams & Participants & Team Size & Tokens & Size   & Repetitions  & Round Dur.  & Study Dur. & Recruitment Platform  \\ \midrule
		TEAMS  & 63   & 252  & 3--4 & 573k      & 110k utterances       & 2          & 30min           & 1.5hr   & Local     \\
		ASIST  & 64   & 192  & 3 & ---      & ---  & 2                                & 15min & 3.5hrs        & Online, Local \\
		CerealBar & N/A & 264 & 2 & 325k        & 24k utterances  & N/A                               & 16.5min  & --- & Crowdworker\\
		PhotoBook & N/A   & 1,514 & 2 &     984k      & 164.6k utterances      & N/A                    & ---           & 14.2m & Crowdworker   \\
		HCRC map task & 32 & 64 & 2 & 150k & 18hrs   & 4                &     ---           &    ---    & School       \\
		PentoRef & 63 & 127 & 2 & 216.3k & 23k utterances & --- & --- & --- &   --- \\
		KTHTangrams & 42 & 84 & 2   &      68k    & 11hrs/15k utterances  &      ---     &     ---        &      15min     & Local    \\
		Cards & N/A & ---  & 2   &  282k         & 45,805 utterances  &   N/A           &       8.5min         & ---             & Crowdworker \\
		\cpsrp{} Pilot      & 8    & 35 & 3--4   &   8k    & 1.5k utterances      & 9           & 4-6min         & 1.5hr     & Local \\
		\bottomrule
	\end{tabular}
 	\caption{Statistics of openly available corpora collected during a CPS task. Repetitions are the number of tasks rounds completed by each team. Study durations are often longer than the time required to complete each round because they include surveys. Local recruitment indicates the local community and can include members beyond the research institution. --- indicates information was not reported. Datasets with crowdworkers did not control for the number of repetitions workers could complete, and teams did not necessarily have unique workers, therefore stats reported are N/A.}
	\label{tab:casestudy_stats}
\end{table*}

%% file: src/tab/example_cps_data.tex
\begin{table*}[ht]

\begin{subtable}[!htp]{0.45\textwidth}\centering
    \scriptsize

    \begin{tabular}{p{.9\linewidth}}\toprule
        \textbf{--- Level 1 Round 1 ---} \\
        Mundert: no slow :(  \\
        Mundert: spam damage?  \\
        oobma: sure  \\
        Mundert: oh wait   \\
        oobma: we got different towers  \\
        Mundert: we have different towers  \\
        TommyVCT: I guess just yolo  it  \\
        omar: yeah   \\
        Mundert: ok mine only do damage  \\
        TommyVCT: I have the one that makes enemies sluggish  \\
        TommyVCT: looks like we got a lot of money  \\
        omar: mine only do damage too   \\
        TommyVCT: oops nevermind we are broke lol  \\
        Mundert: easy win  \\
        oobma: gogo?  \\
        omar: lets go  \\
        TommyVCT: gogogo  \\
        TommyVCT: it's funny that they went backwards  \\
        Mundert: oh it looks like we can kill box with the tree that frightens enimies  \\
        Mundert: and the vine one  \\
        omar: we probably went overboard lol  \\
        Mundert: and area damage would be good with that too  \\
        TommyVCT: ez  \\
        omar: probably should save money next time to get higher score  \\

        \textbf{--- Level 1 Round 2 ---} \\
        Mundert: wait if we lose do we still get a score  \\
        omar: its the same enemies right?   \\
        TommyVCT: looks like it's the same  \\
        omar: lets have the same setup at the start and nothing after   \\
        omar: to save money   \\
        Mundert: ok christmas tree and vine killbox?  \\
        TommyVCT: I got the same roll of the tools too  \\
        Mundert: whatever the cannon was for area damage?  \\
        Mundert: spam em  \\
        omar: who has the cannons?  \\
        oobma: was it the cannon? i only had 1 i thought  \\
        oobma: pretty sure it was the plant thing  \\
        omar: sorry the catapult   \\
        omar: its missing here   \\
        Mundert: cannon does area damage  \\
        TommyVCT: I'll try to deter the enemies using the diamond  \\
        Mundert: so we should use that for a killbox  \\
        Mundert: single target is kinda bad for a killbox  \\
        Mundert: so im not placing my catapults if we do that  \\
        oobma: how many cannons then  \\
        oobma: 4 more?  \\
        omar: maybe 2?  \\
        Mundert: sure  \\
        Mundert: hoewver we can afford and more trees and vines too right  \\
        TommyVCT: wait  \\
        TommyVCT: should I sell my diamonds?  \\
        Mundert: maybe those crossbow things in the line as well  \\
        Mundert: not all  \\
        Mundert: right  \\
        Mundert: because slow is also good  \\
        omar: sell the diamonds in tile (8,9) and (8,8)  \\
        oobma: imo the cross bows would be good at 8,9  \\
        oobma: and 8,8  \\
        omar: ill putt a cross bw there   \\
        Mundert: agree  \\
        TommyVCT: That's all I got  \\
        Mundert: >  \\
        Mundert: ?  \\
        TommyVCT: The tank or controller like thingy is for faster emenies  \\
        Mundert: wait why is the tank there  \\
        omar: but could you sell tile 8,9?  \\
        TommyVCT: oh I put there  \\
        omar: crossbow is better there   \\
        Mundert: agree  \\
        Mundert: aight  \\
        Mundert: nice  \\
        omar: much better   \\
        Mundert: i dont think we need the tank  \\
        TommyVCT: yeah it's kinda useless  \\
        Mundert: more tree and vine and other such area of affect towers  \\

\bottomrule
    \end{tabular}
        \caption{Sample conversation from Level 1.}
    \label{app:cps_exs}
\end{subtable}
\hfill
\begin{subtable}[!htp]{0.45\textwidth}\centering
    \scriptsize

    \begin{tabular}{p{.9\linewidth}}\toprule
<speaker>tjwill</speaker> <chat\_text>Full map ones we probably want bottom left </chat\_text> \\
<action>BUY</action> <tower\_type>DISCOUNT</tower\_type> <location>(10, 0)</location> <user>ManedWlf</user>  \\
<speaker>tjwill</speaker> <chat\_text>If you do a 3x3 grid, empty the center and I'll put an upgrade gem. </chat\_text> \\
<action>BUY</action> <tower\_type>MULTI</tower\_type> <location>(13, 5)</location> <user>schou01</user> \\
<action>BUY</action> <tower\_type>MAP</tower\_type> <location>(0, 14)</location> <user>ManedWlf</user> \\
<action>BUY</action> <tower\_type>MAP</tower\_type> <location>(0, 15)</location> <user>ManedWlf</user> \\
<action>BUY</action> <tower\_type>MAP</tower\_type> <location>(0, 13)</location> <user>ManedWlf</user> \\
<action>BUY</action> <tower\_type>MAP</tower\_type> <location>(1, 13)</location> <user>ManedWlf</user> \\
<speaker>tjwill</speaker> <chat\_text>Then we want a discount tower on the outside, upgrades are Sponsive! </chat\_text> \\
<action>BUY</action> <tower\_type>MAP</tower\_type> <location>(2, 13)</location> <user>ManedWlf</user>  \\
<action>BUY</action> <tower\_type>SUPPORT</tower\_type> <location>(1, 14)</location> <user>tjwill</user>  \\
<action>BUY</action> <tower\_type>MAP</tower\_type> <location>(1, 15)</location> <user>ManedWlf</user>  \\
<action>BUY</action> <tower\_type>MAP</tower\_type> <location>(2, 15)</location> <user>ManedWlf</user>  \\
<action>BUY</action> <tower\_type>MAP</tower\_type> <location>(2, 14)</location> <user>ManedWlf</user>  \\
<action>BUY</action> <tower\_type>MAP</tower\_type> <location>(1, 12)</location> <user>ManedWlf</user>  \\
<speaker>schou01</speaker> <chat\_text>where do we want to focus our offense? </chat\_text> \\

\bottomrule
    \end{tabular}
        \caption{Sample interaction where tjwill suggests placing MAP towers in the bottom left corner of the level in a 3x3 grid, leaving the center empty to place a DISCOUNT tower. ManedWlf proceeds to follow the proposal sending a text message, showing agreement with the proposal through the strategy implementation.}
    \label{app:cps_exs_1}
\end{subtable}
\caption{Example conversations and interactions from our \cpsrp{} pilot study.}
\label{tab:cps-examples}
\end{table*}

%% file: src/fig/money_unspent.tex
\begin{figure}[!thb]\centering
    \includegraphics[width=0.9\linewidth]{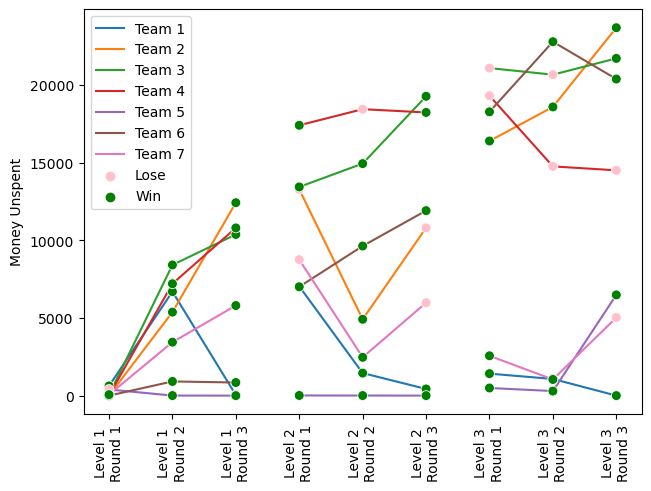}
    \caption{Money remaining for every team, higher is better. The task goal was to minimize expenditures and still win.}
    \label{fig:money-unspent}
\end{figure}

%% file: src/fig/presurvey.tex
\label{app:presurvey}
\begin{figure*}[!ht]
\centering
    \includegraphics[width=.7\textwidth]{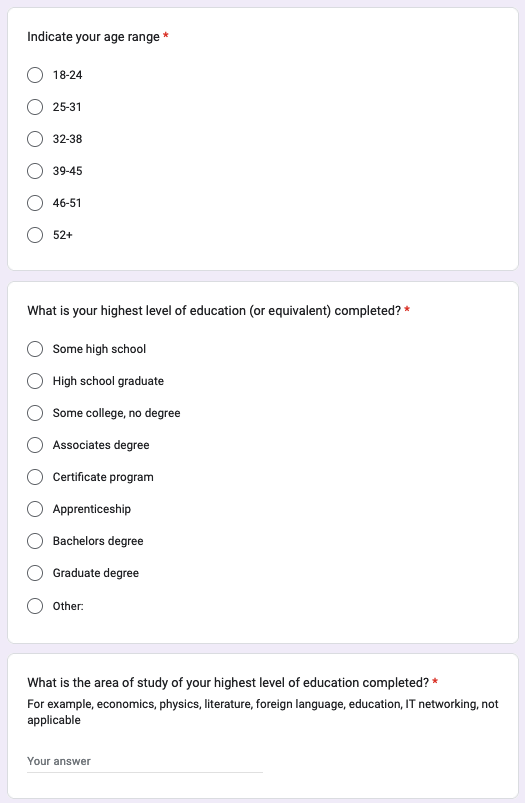}
\end{figure*}
\begin{figure*}[!ht]
\centering
    \includegraphics[width=.7\textwidth]{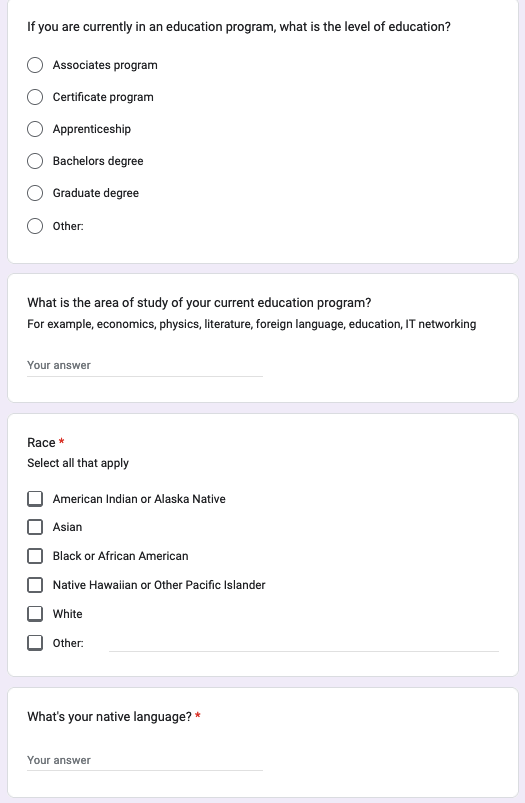}
\end{figure*}
\begin{figure*}[!ht]
\centering
    \includegraphics[width=.7\textwidth]{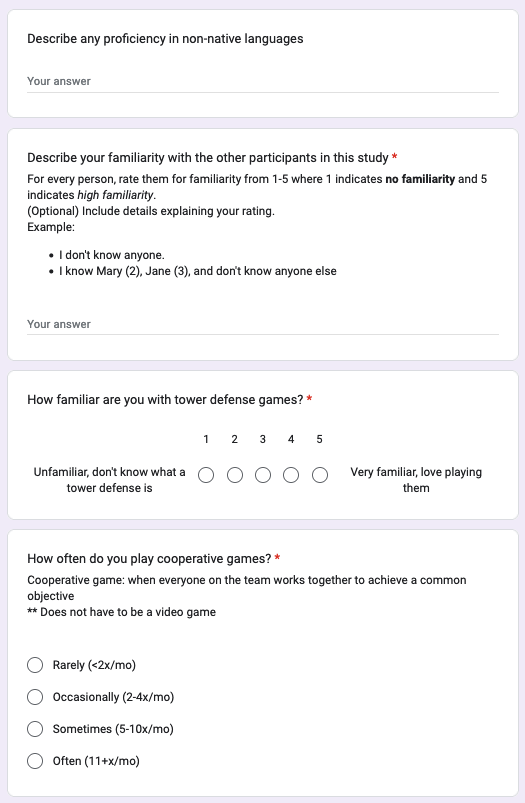}
\end{figure*}